\documentclass{article} 
\usepackage{emulateapj}
\usepackage{color}
\usepackage[]{epsfig}
\hoffset -1.5truecm \lefthead{Menci et al.} 
\righthead{} 
\def\newline{\hfil\break}
\begin{document}
\title{EARLY HIERARCHICAL FORMATION OF MASSIVE GALAXIES TRIGGERED BY INTERACTIONS} 

\author{N. MENCI\altaffilmark{1}, A. CAVALIERE\altaffilmark{2}, 
A. FONTANA\altaffilmark{1}, E. GIALLONGO\altaffilmark{1}, F. POLI\altaffilmark{1}, 
V. VITTORINI\altaffilmark{1}}

\altaffiltext{1}{INAF - Osservatorio Astronomico di Roma,
via di Frascati 33, 00040 Monteporzio, Italy}
\altaffiltext{2}{Dipartimento Fisica, II Universit\`a di Roma, 
via Ricerca Scientifica 1, 00133 Roma, Italy}

\begin{abstract}
To address the problem concerning the early formation of stars in massive galaxies, 
we present the results of a semi-analytic model of galaxy formation which 
includes a physical description of starbursts triggered by galaxy interactions. 
These originate from the 
destabilization of cold galactic gas occurring in galaxy encounters, which in 
part feeds the accretion onto black holes powering quasars, and in part drives 
circumnuclear starsbursts at redshifts $z\approx 2-4$, preferentially in massive 
objects. This speeds up the formation of stars in massive galaxies at 
high redshifts without altering it in low mass galactic halos. 
Thus, at intermediate $z\approx 1.5-2$ we find that a considerable fraction of the stellar 
content of massive galaxies is already in place, at variance with the predictions 
of previous hierarchical models. The resulting high-$z$ star formation rate and 
B-band luminosity functions, and the luminosity and redshift distribution of galaxies 
in K-band at $z\lesssim 2$ are all in good agreement with the existing observations 
concerning the bright galaxy population.
\end{abstract}


\section {Introduction}
Current hierarchical theories of galaxy formation envisage the build up of 
stars in galaxies as a gradual process driven by the continuous growth of the
host dark matter (DM) galactic halos through repeated merging events. 
These progressively increase also the gas mass in the growing galaxies;  
its subsequent radiative cooling is partially counteracted by the 
heating due to the winds from Supernovae originated by parent massive stars. 
The net result of such a ''quiescent'' star formation is a gradual increase 
of the stellar content of the galaxies. 

But when the picture is quantitatively modeled 
through the semi-analytic models (SAMs, Kauffmann et al. 
1993;  Somerville \& Primack 1999; Cole et al. 2000; Menci 
et al. 2002) the resulting amount of stars formed at high $z$ in the {\it bright} 
galaxy population is lower than indicated by several observational results.  


Among these we mention first the excess of the cosmic 
density of bright galaxies observed to be in excess over the 
model predictions at $z\approx 3-5$. True that  several 
groups (e.g., Cole et al. 2000; , Somerville Primack \& 
Faber 2001; Fontana et al. 2003a) 
have shown the canonical SAMs (i.e., with no merging-induced starbursts) to 
provide a total star formation history (integrated over all 
luminosities) consistent with observations. But when one focuses on the UV luminosity density produced 
by the {\it brightest} galaxies only, these SAMs underpredict the luminosity densities observed at 
$z\gtrsim 4$ (Fontana et al. 1999, 2003a). This traces back to the fact that the detailed shape of the  
luminosity functions (LFs) from SAMs are appreciably steeper 
than the observed ones at these redshifts. Actually, the SAMs overpredict the number of faint galaxies 
and, for the canonical SAMs with quiescent star formation only, they 
underpredict the bright ones, see Somerville, Primack \& Faber (2001).

Second, and most important, the SAMs underpredict both the 
bright galaxies observed in the K-band luminosity functions 
at intermediate redshifts ($z\approx 1 - 2$, see Pozzetti et 
al. 2003) and the galaxies brighter 
than $K=20$ counted at $z>1.5$ (Cimatti et al. 2002). 
The emission in the K-band is largely 
contributed by old stellar populations, and so is a measure 
of the amount of stars already formed; thus the above 
observations imply that the star content of massive galaxies 
already in place at $z\approx 2$ is larger than that 
resulting from the gradual star formation typical of 
hierarchical models. Indeed, independent, direct 
observations at $z\approx 2$ of the stellar mass density of 
massive galaxies (in the range $m_*\approx 10^{10}-10^{11}$ 
$M_{\odot}$) yield a fraction close to 0.3 of the present 
value, while the canonical (no-burst) SAMs yield a fraction 
around 0.1 (see Fontana et al. 2003b). The above results 
concur to indicate that the physics of hierarchical galaxy 
formation still lacks some basic process 
enhancing star formation in massive galaxies. 

A candidate for such an enhancement is
constituted by starbursts triggered by galaxy interactions, as proposed by 
Somerville, Primack \& Faber (2001), see also Cavaliere \& Menci (1993). The former  
authors assumed that satellite galaxies contained in the same host DM halo (i.e.,  
in groups or clusters of galaxies) may undergo binary aggregations, which would 
not only affect their mass distribution but also brighten them by triggering 
starbursts. 
In such a model, only encounters leading to outright merging are 
considered; in addition, the 
fraction of galactic gas converted into stars during each burst is taken 
to be constant in time, with a parametrized dependence on the merger mass ratio
which favors major mergers. The resulting starbursts, while improving the match 
with the observed UV LFs at $z=3$ and $z=4$,  
still do not provide enough high-$z$ star formation to account for the 
number of bright ($K>20$) galaxies at $z>1.5$ observed by Cimatti et al. (2002). 

Thus one main problem with the current SAMs is their failure 
to form enough stars at high-$z$ in {\it bright} galaxies, 
so as to match the observed bright end of the $K$-band LFs 
at intermediate $z$. 

Here we address the issue on the basis of a more complete description of the starbursts triggered by 
galaxy encounters. This is based on the physical model for the destabilization of cold 
galactic gas during galactic merging and fly-by developed by Cavaliere \& 
Vittorini (2000, CV00). The destabilized gas is assumed to feed in part the 
accretion onto a central super-massive black hole (BH) so powering 
quasar-like emission, and in part a burst of  star formation. The amount of 
destabilized gas, the duration of the bursts, and their rate set by the galaxy 
encounters are determined by the physical properties of the galaxies and of 
their host halos, groups or clusters. 
While CV00 adopted simplified derivations for the galactic properties 
in common DM halos, here such properties are self-consistently computed from the SAM 
presented in Menci et al. (2002). 

In a previous paper 
(Menci et al. 2003) we have shown that the quasar (QSO) evolution 
resulting from such a model is in excellent agreement with a whole set of 
observables. Here we focus on the circumnuclear starbursts originating from the 
fraction of cold gas destabilized in galaxy encounters 
complementary to the gas accreted to the central supermassive BH;  
such starbursts mainly affect the massive galaxy population.

We recall in Sect. 2 the basic features of the SAM presented in Menci 
et al. (2002), while in Sect. 3 we present our treatment of the new processes included in 
our present model, i.e., 
the fly-by events and the associated destabilization of cold galactic gas. The results are presented in 
Sect. 4, while Sect. 5 is devoted to conclusions and  discussion.  

\section{Modeling the Galaxy Evolution}

To describe the galaxy evolution in the hierarchical scenario, we adopt the SAM 
described in detail in Menci et al. (2002); here we recall the basic points. 
We consider both the host DM halos containing the galaxies
(i.e., groups and clusters of galaxies with mass $M$, virial radius $R$ and circular velocity $V$), 
and the DM clumps (with mass $m$, radius $r$ and 
circular velocity $v$) associated with the 
individual member galaxies. The former grow hierarchically to larger sizes 
through repeated merging events (at the rate given in Lacey \& Cole 1993), while 
the latter may coalesce either with the central galaxy in the common halo due to 
dynamical friction, or with other satellite galaxies through binary 
aggregations. 
The timescale for the dynamical friction and the binary merging 
processes, and so the probability for such processes to occur in each timestep, 
are given by eq. (2) and (4) in Menci et al. (2002). 

We assume initially (at $z\approx 10$) one galaxy in each host structure, with the 
latter following the Press \& Schechter mass distribution. 
The probability for the merging processes (dynamical friction and binary aggregations) 
to occur during the hierarchical growth of the hosting structure yields  
the differential distribution function $N(v,V,t)$ of 
galaxies with given $v$ in hosts with circular velocity $V$ at the cosmic time $t$.   
From $N(v,V,t)$ we derive the number $N_T(V,t)$ of galaxies in a host 
halo (membership), and the overall distribution of galaxy circular velocity 
$N(v,t)$ irrespective of the host. 

The properties of the baryons (gas and stars) contained in the galactic DM clumps 
are computed as follows. Starting from an initial amount $m\,\Omega_b/\Omega$ of gas 
at the virial temperature of the galactic halos, we compute the mass $m_c$ of cold baryons 
within the cooling radius. 
The disk associated to the cold baryons will have a radius $r_d(v)$,  
rotation velocity $v_d(v)$, and dynamical time $t_d = r_d/v_d$, 
all computed after Mo, Mao \& White (1998). 
From such a cold phase, stars are allowed to form at the rate 
\begin{equation}
\dot m_* = {m_c\over t_{dyn}}\,\Big({v\over 200\,{\rm km\,s^{-1}} }\Big)^{-\alpha_*}
\end{equation}

Finally, a mass $\Delta m_h=\beta\,m_*$ is returned 
from the cool to the hot gas phase due to the energy fed back   
by canonical type II Supernovae associated to $m_*$; the feedback efficiency is taken to be 
$\beta= (v/v_h)^{\alpha_h}$. The values adopted for the parameters 
$\alpha_*=-1.5$, $\alpha_h=2$ and $v_h=150$ km/s  
fit both the local B-band galaxy LF and the Tully-Fisher relation, 
as illustrated by Menci et al. (2002). 

At each merging event, the masses of the different baryonic phases are refueled 
by those in the merging partner. The further increments $\Delta m_c$, $\Delta 
m_*$, $\Delta m_h$ from cooling, star formation and feedback are recomputed on 
iterating the procedure described above. 

Thus, for each galactic $v$, the star formation defined by eq. (1) is 
driven by the cooling rate of the hot gas and by the rate of refueling of cold gas, 
which in turn is related to the progressive growth of the total galactic mass along 
the merging tree. The related brightening of galaxies is a gradual process, 
and the associated star formation history is often referred to as ''quiescent'' star
formation. 

The integrated stellar emission 
$S_{\lambda}(v,t)$ at the wavelength $\lambda$ is computed by convolving the 
SFR with the spectral energy distribution $\phi_{\lambda}$ obtained from population 
synthesis models: 
\begin{equation}\label{sed}
S_{\lambda}(v,t) = \int_0^t\,dt'\,\phi_{\lambda}(t-t')\,\dot m_*(v,t')~. 
\end{equation}
In the following we adopt $\phi_{\lambda}$ taken from Bruzual \& Charlot (1993), 
with a Salpeter IMF. The dust extinction is computed as described in Menci et al. (2002). 

All computations are made in a $\Lambda$-CDM cosmology with 
$\Omega_0=0.3$, $\Omega_{\lambda}=0.7$, a baryon fraction 
$\Omega_b=0.03$, and Hubble constant $h=0.7$ in units of 100 
km s$^{-1}$ Mpc$^{-1}$. 

\section{Star Formation triggered by Galaxy Encounters}
In the present work, we upgrade the above model by adding a treatment of 
fly-by events (i.e., encounters which do not lead to bound merging) and of the 
related bursts of star formation. 
Indeed, galaxy encounters are expected to destabilize part of the available cool gas by 
causing it to loose or transfer angular momentum (Barnes \& Hernquist 1998; 
Mihos 1999); this triggers gas inflow. The gas funneled inward may end up 
in accretion onto a central supermassive BHs, but also in a nuclear starburst 
(see Sanders \& Mirabel 1996). A quantitative model to derive the fraction $f$ of cold 
gas destabilized by the encounters has been worked out by CV00; 
here we recall the guidelines in terms suitable for direct implementation in our SAM. 

The fraction of cold gas which is destabilized in each interaction event and feeds 
the starbursts is derived from eq. A3 of CV00 in terms of the  variation 
$\Delta j$ of the specific angular momentum $j\approx Gm/v_d$ of 
the gas; in slow, grazing encounters between galaxies with relative velocity $V_r$
one obtains: 
\begin{equation}
f(v,V)\approx {3\over 8}\,
\Big|{\Delta j\over j}\Big|=
{3\over 8}\Big\langle {m'\over m}\,{r_d\over b}\,{v_d\over V_r 
 }\Big\rangle\, . 
\end{equation}
The effective impact parameter $b=max[r_d,{\overline{d}(V)}]$ is   
the maximum between the radius $r_d$ and the average distance 
${\overline{d}}(V)=R/N_T^{1/3}$ of the galaxies in the halo; 
we take $V_r$ to be twice the one-dimensional velocity dispersion 
$\sigma_V=V/\sqrt{2}$ of the host halo. 
The first approximate equality is derived by considering the amount of gas 
which is in centrifugal equilibrium outside a circumnuclear region of fixed size; 
since the extension $\Delta r$ of the outer region is proportional to $j$ (see, e.g., Mo, Mao 
\& White 1998), a loss of angular momentum implies a negative $\Delta r$ and 
hence a mass flow toward the circumnuclear region. As for the 
prefactor, it accounts for the probability 1/2 of inflow rather than outflow 
related to the sign of $\Delta j$. We assume that $1/4$ of the inflow feeds the 
central BH, while the remaining fraction is assumed to kindle circumnuclear 
starbursts, see Sanders \& Mirabel (1996); thus, the starbursts efficiency 
$f$ in eq. (3) is three times larger than the complementary BH accretion efficiency 
shown in fig. 1 of Menci et al. (2003). 

The last equality in eq. (3) has been derived by CV00 by computing the variation $\Delta j= 
Gm'r_d/Vb$ of the angular momentum in the galaxy from the gravitational torque 
(proportional to the disk size $r_d$) 
exerted by the partner galaxy with mass $m'$; this is time-integrated along the partner 
orbit. Note that the dependence of $\Delta j$ on $r_d$ causes 
the the amount of destabilized gas to be larger in more massive systems. 
The average runs over the probability of finding a galaxy with mass $m'$ 
in the same halo $V$ where the galaxy $m$ is located. 

When inserted into our SAM and applied to the BH accretion and to the related 
QSO emission (Menci et al. 2003), the above model 
has proven to be very successful in reproducing the observed properties 
of the QSO population from $z=6$ to the present, including the rapid decline of their
densities from $z\approx 2.5$ to $z=0$ and even the detailed changes of their LF 
from $z\approx 5$ to $z=0$. 
In the present picture, the interaction-driven starbursts constitute the 
''counterpart'' of the BH accretion powering the QSOs. 

To compute the average effects of the cold gas destabilization 
on the star formation rate and hence on the galactic emission, we 
derive the probability for a given galaxy to be in a burst phase. This is 
defined as the ratio $\tau_e/\tau_r$ of the duration of the burst to the 
average time interval between bursts. 

In turn, 
such quantities can be derived from a detailed analysis of 
the orbital parameters (see Saslaw 1985); following CV00,  
here we just recall that tides effective for angular 
momentum transfer (as considered in eq. 3) require two 
conditions: the interaction time is to be comparable with the 
internal oscillation time in the galaxies involved (resonance); 
the orbital specific 
energy of the partners is not to exceed the sum of the specific internal 
gravitational energies of the partners. The rate of such 
encounters $\tau_r^{-1}$ is then given by Saslaw (1985) in 
terms of the distance $r_t\approx 2r$; for a galaxy with 
given $v$ inside a host halo with circular velocity $V$ the result 
reads
\begin{equation} 
\tau_r^{-1}=n_T(V)\,\langle \Sigma(v,V)\rangle \,V_{r}(V)~, 
\end{equation}
where $n_T=3\,N_T/ 4 \pi$$R^3$, and the 
cross section is averaged over all partners with effective tidal radius $r'_t$ 
in the same halo $V$. The membership $N_T$, and the distributions of $v'$, 
$r_t'$ and $V_{r}$ are computed from the SAM described in \S 2. 

The condition for a grazing encounter defines the encounter duration
$\tau_{e}= \langle (r_t+r_t^{'})/V_r\rangle$ (with the upper limit given by 
$\tau_r$ in eq. 4) where the average is again over all partners with tidal radius 
$r'_t$ in the same host halo $V$. 
It must be noted that the cross section in eq. (4)
determines the probability for any grazing encounter including the fly-by events, 
which are in fact more frequent than major mergers and dominate the statistics
of galaxy encounters. We also stress that in our model the eq. (4) determines only the probability 
$\tau_e/\tau_r$ of finding a galaxy in the burst phase, but 
does not affect the evolution of the galaxy mass function. 
This is instead determined by the processes of dynamical friction and binary aggregation proper, 
for which we retain the cross sections given in Menci et al. (2002), and  
recalled in Sect. 3. 

The {\it average} SFR associated to the destabilized cold gas during an encounter lasting 
a time $\tau_{e}$ reads, 
\begin{equation}
\Delta \dot m_{*}(v,z)=
\Big\langle
{f(v,V)\,\,m_c(v)\over \tau_{e}(v,V)} 
\Big\rangle_{\tau_e/\tau_r} .
\end{equation}

The average is here over all host halos with 
circular velocity $V$, and is weighted with the probability   
$\tau_{e}/\tau_{r}$ of finding the galaxy in the burst phase.  

We remark that our model includes the effects of both fly-by events and bound 
mergings. Although the former induce starbursts with a lower efficiency 
($f\approx 0.1-0.4$ at $z\gtrsim 3$) they contribute appreciably to the {\it 
average} star formation rate in bursts (eq. 5).  In fact, the fly-by events are 
more probable than major mergers, and hence   provide larger values for the 
probability $\tau_{e}/\tau_{r}$ entering the average in eq. 5. This holds even 
though for major merging events (requiring small relative velocities and hence 
small values of $V_r$ in eq. 3) the efficiency may attain values $f\approx 0.7$, 
consistent with the values 0.65-0.8 obtained in the hydrodynamical simulations 
by Mihos \& Hernquist (1996), and  close to the values observed in SCUBA sources 
and in the local Ultra Luminous Infrared Galaxies (see Blain et al. 2002 and 
references therein; Sanders \& Mirabel 1996). At $z\gtrsim 3$, the contributions 
to eq. (5) from merging and from fly-by events are comparable. 

The {\it average} contribution to the stellar emission $S(v,t)$ from bursts in a galaxy 
with circular velocity $v$ at the time $t$ is given by eq. (2), with 
the quiescent star formation rate $\dot m_*$ replaced by the starbursts rate $\Delta 
\dot m_{*}(v,t)$ given by eq. (5). From $S(v,t)$ we compute the LF given by $N(v,t)\,dv/dS$. 

\section{Results}

In fig. 1 (bottom panel) we show 
the effect of bursts on the B-band LF of galaxies at $z=0$, 
and on the UV LFs $z=3$ and at $z=4$. 

\begin{center}
\vspace{-1.9cm} 
\scalebox{0.45}[0.45]{\rotatebox{0}{\includegraphics{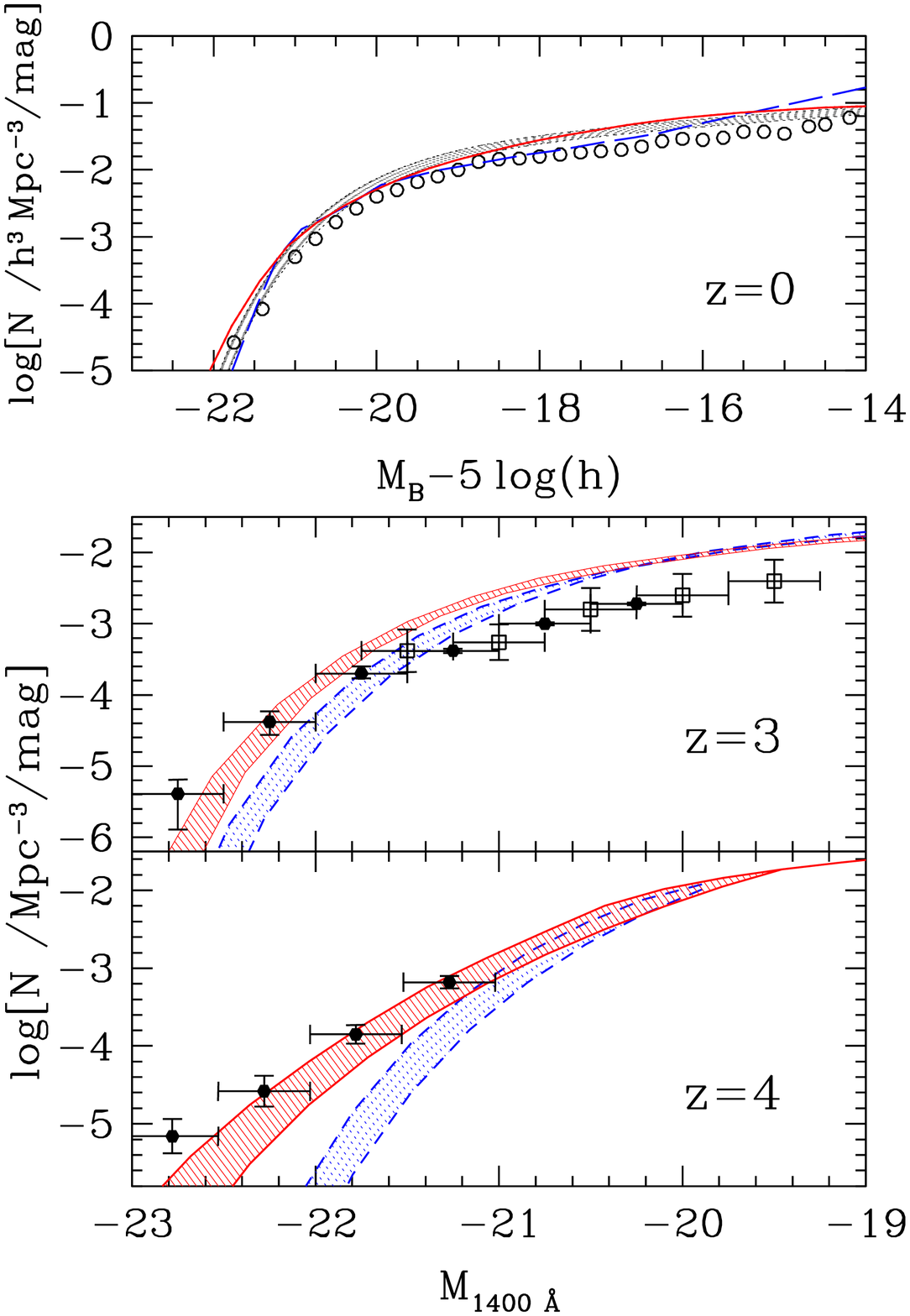}}}
\end{center}
{\footnotesize
\vspace{-0.25cm } 
Fig. 1. - The predicted LFs of galaxies in the quiescent (blue dashed lines) and 
starburst modes (red solid lines)  at $z\approx 0$ 
(in the B-band, top panel), and at $z\approx 3$ and $z\approx 4$ (at {1400 \AA}, middle and 
bottom panels, respectively). 
In the top panel, the shaded area corresponds 
to the LF measured by the Sloan Digital Sky Survey 
(Blanton et al. 2000), and the circles to the data from the 2dFGRS 
survey (Madgwick et al. 2002).
In the middle and bottom panels, the two red solid lines refer to the burst model and 
bracket the uncertain due to different dust extinction curves (represented by the 
shaded regions) for given 
dust-to-gas ratio, see text. Similarly for the blue dashed curves (and the comprised 
shaded areas) referring to the 
quiescent model. The spectroscopic (solid squares) and the photometric (open squares) 
data are from Steidel et al. (1999).  
\vspace{0.2cm}}

As the emission in the UV (and, to a minor extent, that in the B-band) is 
contributed by massive, short-living stars, the LFs in 
fig. 1 show the effect of bursts on the instantaneous star formation 
at low and high redshifts. 
At low $z$ (top panel) the LFs are little affected by starbursts whose 
effect, if anything, is to make the model LF closer to that  
measured by in Sloan survey (Blanton et al. 2000). 

At high $z$, the predictions for the UV LFs are shown in the bottom panels, and 
compared with data uncorrected for extinction, since the dust absorption is 
included in the model. In particular, we implemented in our SAM the Milky Way, 
the Small Magellanic Cloud (SMC) and the Calzetti extinction curves, with the 
dust optical depth fixed by the fit to the local LF.  The related uncertainties 
in the model predictions are illustrated by showing (for both the quiescent and 
the starburst modes) the brightest and the faintest LFs (corresponding to the 
SMC and to the Calzetti curves, respectively). Note that at high $z$ the 
interactions are more effective in stimulating starbursts; at $z=4$, compared with the 
quiescent mode, they brighten the galaxy LF by an amount which grows with the 
luminosity to reach about $0.7$ mag in the brightest magnitude bin.  Note also that 
the inclusion of bursts does not appreciably affect the LFs for galaxies fainter 
than $M_B\gtrsim -20$. 

Indeed, burst are more efficient in high mass than in low mass systems, since 
the former have larger cross section for encounters (see eq. 4). In addition, 
in each encounter the amount of destabilized gas is larger in more massive 
galaxies, since the loss of angular momentum is larger for larger disks  
as discussed below eq. 3. 

As a function of time, fig. 1 shows that galactic encounters are increasingly effective 
in stimulating starbursts for increasing $z$. In fact, the interaction rate 
$\tau_r^{-1}$ and the accreted fraction $f$ decrease with increasing $t$, since 
in the growing host halos the growing membership $N_T(V)$ is offset by $R$, 
$V$ and $V_{r}$ increasing (see eqs. 3, 4). In a group,
the {\it average} values for $f$ drop from 
from a few $10^{-1}$ to several $10^{-3}$ on going from $z\approx 3$ to $z=0$, 
as shown in fig. 1 of Menci et al. 
(2003) (the latter paper shows the efficiency for BH accretion which is three times lower 
than the efficiency for starbursts, as discussed below eq. 3); 
higher values up to $f\gtrsim 0.4$ are instead attained at $z\gtrsim 3$.

Thus, at low redshifts the lower densities and larger relative velocities of 
the galaxies contained in the same host environment suppress the 
effectiveness of the interactions and hence of the associated starbursts. 

\begin{center}
\vspace{-0.1cm} 
\scalebox{0.43}[0.42]{\rotatebox{0}{\includegraphics{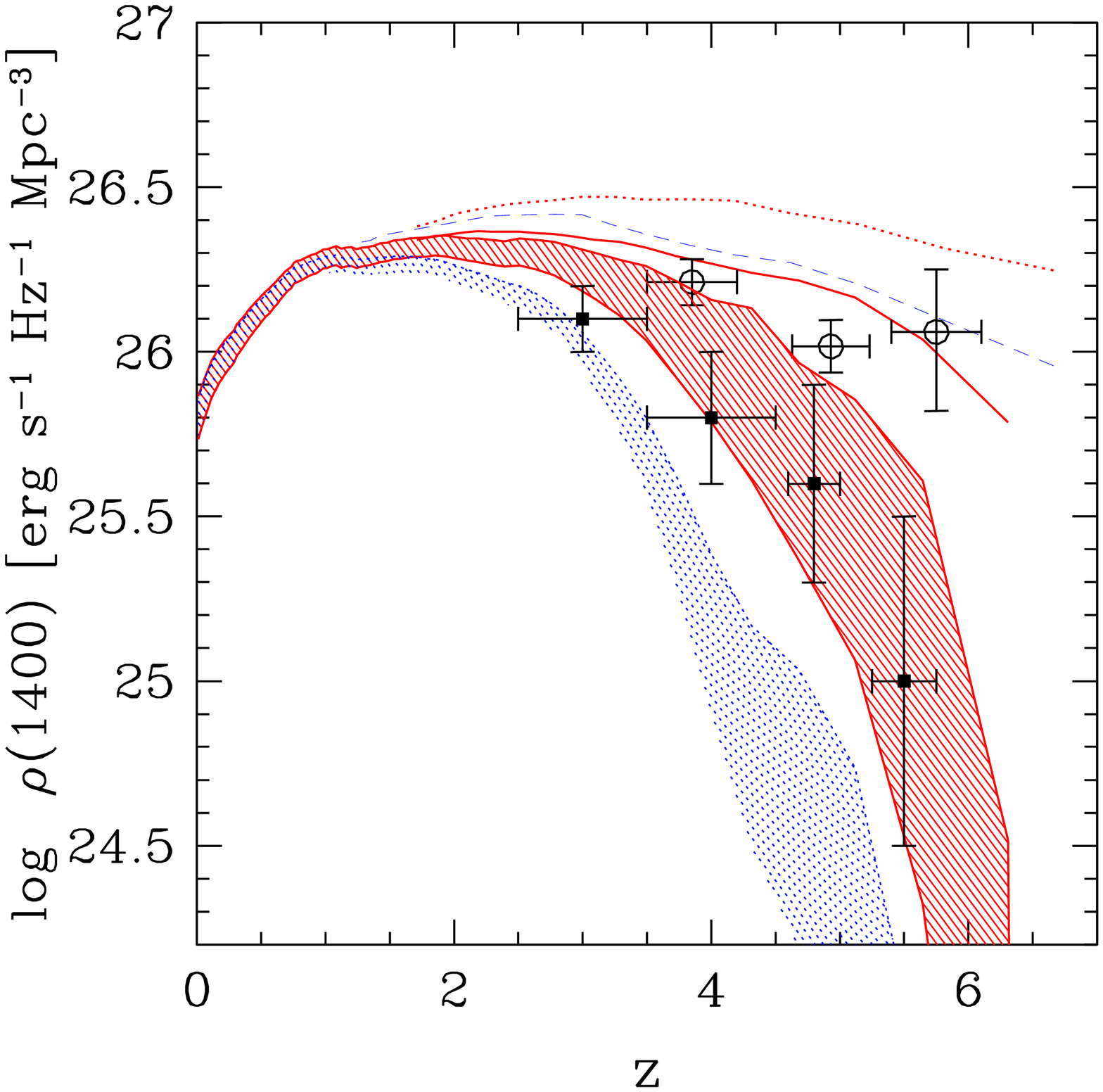}}}
\end{center}
{\footnotesize
\vspace{-0.25cm } 
Fig. 2. - The predicted UV luminosity densities (at {1400 \AA}) compared with data.  
Solid squares refer to the data for galaxies with $m_Z<25$ (Fontana et al. 2003a), 
while the shaded areas (heavy and red for the model with bursts, light and blue for the pure 
quiescent model) represent the model predictions -- at the same limiting magnitude -- 
adopting different extinction curves in the models (see text). 
In addition we plot the UV luminosity density 
obtained by Giavalisco et al. (2003, empty circles) by integrating the observed UV LFs
down to $0.2\,L_{*3}$ (where $L_{*3}$ is the characteristic luminosity of 
observed Ly-break galaxies at $z=3$), compared with the prediction of our model 
for the same luminosity cut (red heavy solid line) averaged over the three extinction 
curves we considered in the text. 
We also show the total UV luminosity density in the burst (red dotted line) and 
quiescent (blue dashed line) models, averaged over the dust extinction as above. 
\vspace{0.2cm}}

Such a picture is borne out by fig. 2, where we show the effect of 
bursts on the cosmic UV luminosity density produced by galaxies, and compare it 
with existing observations. These are based on HST and VLT imaging surveys aimed at 
identifying high-$z$ galaxies. The data concerning the UV luminosity density contributed 
by the bright ($m_Z<25$) galaxy population are compared 
with the model predictions; for these we also show the uncertainties 
due to the different extinction curves, represented as shaded areas. 
In addition, we compare with the recent data from 
Giavalisco et al. (2003) concerning the density contributed 
also by fainter galaxies, with the lower limit in luminosity 
provided in the caption. 

We note that when the UV luminosity density contributed only by bright 
($m_Z<25$) galaxies is considered, the canonical SAMs (with no bursts) 
underpredict the observed values. This is because such models fail to provide 
enough star formation to match (once  dust is included) the bright end of the UV 
LFs at $z\gtrsim 4$ (see fig. 1). However, when one considers the total UV 
luminosity density (obtained by integrating the LFs over all luminosities) the 
same models provide a sustained luminosity density (the dashed line) up to large 
$z$. This is because the shape of the LFs from such SAMs are appreciably steeper 
than the observed ones at high $z$; while such models underpredict the density 
of bright sources, they overpredict the number of faint galaxies. The two 
effects balance as to yield a sustained total UV luminosity 
density at high $z$, as shown in fig. 2. 

On the other hand, the burst model naturally provides more 
star formation in massive galaxies and so matches in detail the 
observed UV luminosity density from bright galaxies with 
$m_Z<25$. We also compare our burst model with the observed UV 
luminosity density contributed by fainter galaxies. Following  
the discussion above, any such comparison depends critically on 
the lower luminosity limit for the integration of the UV 
LFs. So in comparing our prediction (heavy solid line in 
fig. 2) with the data from Giavalisco et al. (2003) we adopt 
the same luminosity cut used by such authors (see caption), 
finding a good agreement with observations. 

As a result, in the starburst scenario the fraction  of stars predicted to be 
already formed in {\it massive} galaxies by $z\approx 2$ is significantly larger 
then in the quiescent models. This, in turn affects the galaxy LFs and the 
corresponding redshift distributions in the $K$ band; here the emission is 
largely contributed by evolved stellar populations, so probing the total amount 
of stars which have been assembled in galaxies by a given cosmic time. 
In the following, we shall focus on such issues since, as we recalled in Sect. 1, 
matching the K-band observables and the stellar mass density at intermediate redshifts 
constitutes one main problem for canonical SAMs. 

In fig. 3 we compare the results with observations
obtained from the K20 survey (Cimatti et al. 2002; Pozzetti et al. 2003). 
The starbursts brighten the LF by $\sim 0.5$ mag at $z\approx 1.5$, so 
matching the observed shape of the LFs (top panel).  Meanwhile the faint end of the LFs 
is left almost unchanged. 
This is again a consequence of the larger effectiveness of the bursts in more massive 
galaxies, which is due to the physics of interaction-driven 
bursts combining with the statistics of encounters. On the one hand, in each bursts 
galaxies with larger disk sizes undergo larger losses of angular momentum as a 
consequence of larger gravitational torques, as explained below eq. (3); on the 
other hand, larger galactic sizes favor the encounters, as 
described in detail below eq. (4).  
Correspondingly, the burst model {\it matches} the observed number of luminous 
($m_K<20$) galaxies at $z\gtrsim 1.5$, while the quiescent model underpredicts 
the number by a factor $\sim 3-4$ (see bottom panel in fig. 3). 

\begin{center}
\vspace{-0.2cm} 
\scalebox{0.45}[0.45]{\rotatebox{0}{\includegraphics{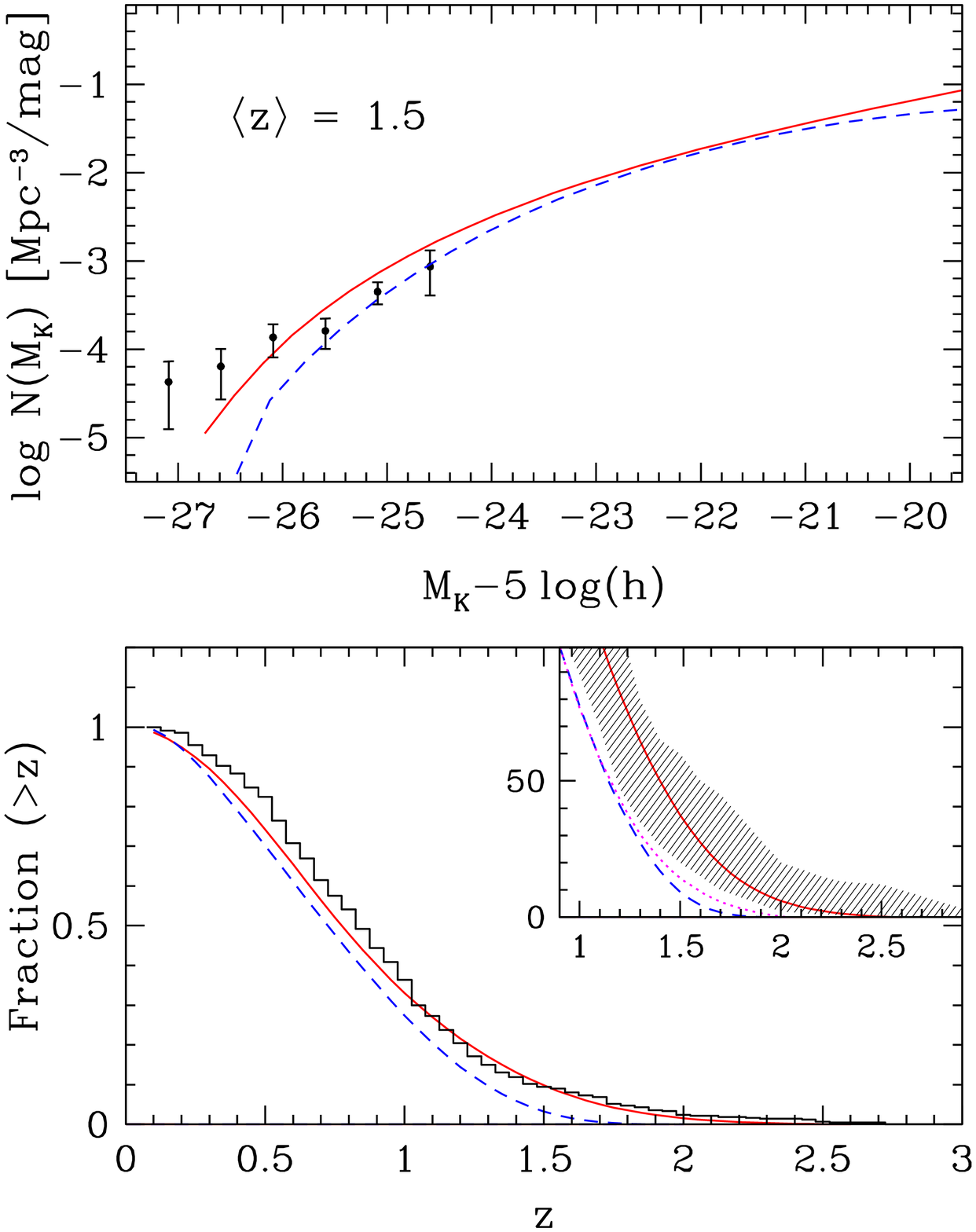}}}
\end{center}
{\footnotesize
\vspace{-0.25cm } 
Fig. 3. - Upper panel: the K-band galaxy LFs at $z=1.5$ 
in the quiescent (blue dashed line) and 
in the starburst model (red solid line) are compared with the data from 
the K20 survey (Pozzetti et al. 2003).
Bottom panel: the corresponding cumulative $z$-distribution of 
$m_K<20$ galaxies are compared with observations from the K20 survey (Cimatti et al. 2002). 
The inset shows 
in detail the cumulative distribution in the range $1<z<3$ with the 
$3\sigma$ Poissonian confidence region (shaded area); the dotted line reproduces from Cimatti et al. 
(2002) the prediction of the SAM by Somerville, Primack, \& Faber (2001). 
\vspace{0.2cm}}

Inspection of fig. 3 shows that our burst model provides a better fit to the 
observed $z$-distribution of galaxies with $m_K<20$ than the SAM by Somerville, 
Primack e Faber (2001), which also included starbursts. The reason is that our 
model provides a higher star formation rate at high redshift. This is shown by 
the dotted line in fig. 2; when the total UV luminosity density of our model is 
converted to a total density of star formation, the resulting star formation 
rate is larger than the Somerville, Primack \& Faber (2001) rate at $z>4$ for 
any reasonable dust extinction adopted in the conversion. The improvement is 
achieved despite of an {\it average} value of the burst efficiency $f$ somewhat lower 
than adopted by Somerville, Primack e Faber (2001). 

Our improvement is due to two circumstances. 
First, the above authors associate starbursts only to bound mergers; for such 
events (requiring very slow encounters with small $V_r$) we obtain from eq. (3) 
values $f\approx 0.7$ comparable to theirs, as these particular interactions are 
maximally effective in inducing loss of gas angular momentum. However, in our 
model we also consider the effect of the more frequent fly-by events not leading 
to outright merging of the involved galaxies. These produce starbursts with a 
lower efficiency $f\approx 0.1-0.4$ (at $z\gtrsim 3$), but dominate the encounter statistics. 
While they yield a low average $f$, the key quantity for $\Delta m_*$ is instead 
the product $f\,\tau_r^{-1}$ (i.e., the efficiency weighted with the interaction 
rate)  which is enhanced by the fly-by events, see eq. (5) and the discussion 
below it. 
Second, the dynamics of sub-halo mergers in our model slightly favors the formation 
of massive galaxies at high-$z$ compared to SPF. This is because 
the cross section we adopt for bound mergers (taken from Menci et al. 2002, eq. 4) is somewhat 
larger than that used by 
Somerville, Primack \& Faber (2001), especially at high $z$. In fact, the 
above authors adopt the cross section derived by Makino \& Hut (1997) from N-
body simulations, strictly valid for encounters between equal galaxies in the 
limit of large relative velocities. Our cross 
section, although reducing to the Makino \& Hut's (1997) in the proper limit of 
merging between equal galaxies with large $V_r$ (as shown in Menci et al. 2002), 
also holds for lower relative velocities. 
This is particularly relevant at high $z\gtrsim 4$, when 
galaxies reside in environments with low velocity dispersions. 

\begin{center}
\vspace{-0.1cm} 
\scalebox{0.45}[0.45]{\rotatebox{0}{\includegraphics{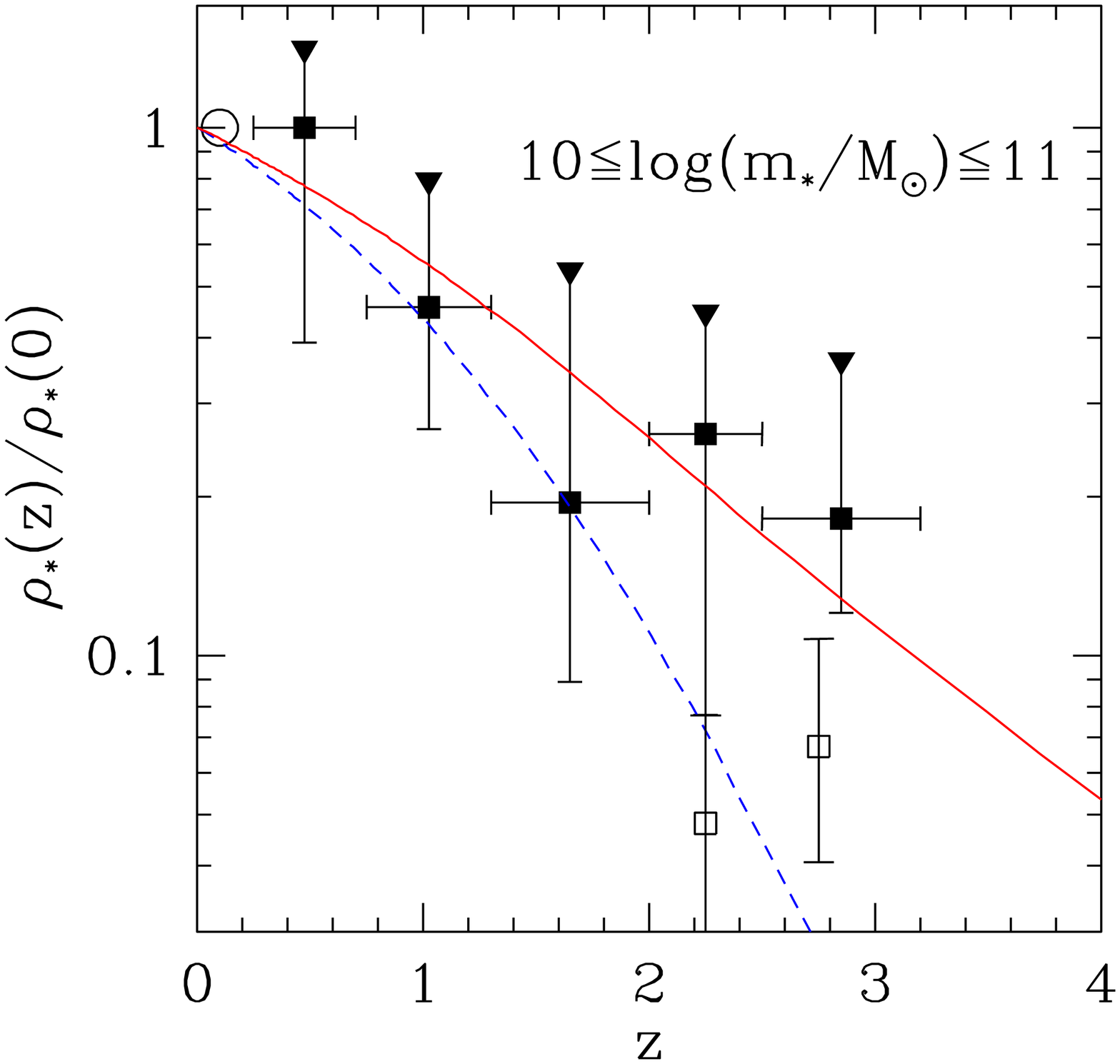}}}
\end{center}
{\footnotesize
\vspace{-0.25cm } 
Fig. 4. - The evolution of the stellar mass density (normalized to the present) 
for galaxies with stellar mass in the range $10^{10}\,M_{\odot}<m_*<10^{11}\,M_{\odot}$
in the starburst mode (red solid line) and in the quiescent mode (blue dashed line) 
is compared with data from Dickinson et al. (2003, open squares) and from Fontana et al. 
(2003, solid squares). The triangles show the upper limit obtained on adopting 
the ''maximal mass'' algorithm of Fontana et al. (2003b). 
\vspace{0.2cm}}

The enhancement in the fraction of stars formed in massive galaxies at high redshift $z>3$ 
due to our treatment of the bursts is illustrated by fig. (4). 
There we show for both the 
burst and the quiescent modes the evolution of the stellar 
mass density in massive ($10^{10}\,M_{\odot}<m_*<10^{11}\,M_{\odot}$) systems, 
and compare it with various observations. 
Although the present data are not sufficient by 
themselves to discriminate between the two models, the plot 
clearly shows that at $z\approx 2$ the model with bursts predicts a fraction of 
stars formed in massive systems which is 2.5 times larger than predicted in the quiescent 
models. 

\section{Conclusions}

We have presented the results of a SAM which includes a physical model for 
starbursts. The latter originate from the destabilization of cold galactic gas 
occurring in galaxy encounters as described by Cavaliere \& Vittorini 
(2000). In this picture, part of the  destabilized gas feeds the accretion onto 
BHs powering the QSOs (see Menci et al. 2003), while the complementary, larger  
fraction produces circumnuclear  starsbursts at redshifts $z\approx 2-4$, 
preferentially in massive objects. This speeds  up the star formation in 
massive galaxies (favored by their larger cross sections for encounters) at 
high redshifts (when the higher densities and the slower relative velocities of 
galaxies in common halos favor strong  interactions). The LFs resulting 
in our model match those of {\it luminous} ($M_B<-21$) Lyman-break galaxies 
observed at $z=3$ and 4, while leaving almost unchanged the model 
predictions for fainter objects. Thus, 
at $z\approx 2$, a larger fraction of the stellar content of the massive 
galaxies is already {\it in place}, at variance with the 
predictions of previous hierarchical models; the resulting K-band LFs and $z$-
distributions match the existing data. 

Compared to Somerville, Primack, \& Faber (2001) our model gives qualitatively similar results, 
but a better fit (see fig. 3) to 
the observed $z$-distribution of bright $K>20$ galaxies at $z\gtrsim 1.5$.  
Such a larger rate is mainly due to the novel inclusion of fly-by events 
which, although producing starbursts with lower 
efficiency ($f\sim 0.1-0.2$ at $z\approx 3$), are more frequent than the bound 
merging events. In our picture, the UV luminosity, and hence the 
star formation rate, of massive galaxies at high redshift
is in part contributed by the few powerful starbursts occurring 
in major merging events, and in part by a more widespread, though less powerful, 
brightening of galaxies almost continuously stimulated by the frequent encounters. 
In this regime, a considerable fraction close to $0.25$ of the present stellar content of massive 
($10^{10}\,M_{\odot}<m_*<10^{11}\,M_{\odot}$) galaxies is formed by $z=2$, see fig. 4. 
As a result, at lower redshifts $z\approx 1.5$ a 
considerable fraction of massive galaxies contains a stellar 
population evolved enough to provide the bright $K$-band 
luminosities required to match the observed LFs and counts.  

In sum, we have shown that making early massive galaxies takes not only 
the early collapse of large DM halos favored in the $\Lambda$-CDM cosmology, 
but also baryons forced to shine starlight by galaxy interactions.  

We stress that our results are achieved on the basis of a physical -- rather than 
phenomenological -- model to derive the burst rate and the amount of cold gas 
converted into stars during the bursts. Moreover, our model naturally connects 
with the accretion onto BHs and with the corresponding QSOs emission. Such a 
unified description is thus able to {\it link} the independent observations 
concerning the luminosity distributions of QSOs and of the galaxies. 

For $z<2$ the merging and the encounter rates decline, and no longer affect the 
evolution of the cold gas reservoirs of massive galaxies.  So, these galaxies 
specifically enter a phase of nearly passive evolution, with redder colors which 
in some cases -- whose probability we shall give in detail elsewhere --
correspond to those of many observed extremely red objects (EROs see, e.g., Daddi   
et al. 2002 and references therein). 
In our model, the beginning and the development of such a phase is 
naturally matched to the dramatic drop of the QSO luminosities; these are 
entirely triggered by the gas destabilized in the encounters, and so are particularly 
sensitive to the current merging and interaction rates of their host galaxies. 

\acknowledgments We thank the referee for useful comments which  
helped to improve our presentation.

\end{document}